# Probing and Tuning Strain-localized Exciton Emission in 2D Material Bubbles at Room Temperature

*Junze Zhou[1], John Thomas[1], Thomas P. Darlington[1], Edward S. Barnard[1], Atsushi Taguchi[2], Adam Schwartzberg[1], Alexander Weber-Bargioni[1]*

[1]*The Molecular Foundry, Lawrence Berkeley National Laboratory, 1 Cyclotron Road, Berkeley, California 94720, USA*

[2]*Research Institute for Electronic Science, Hokkaido University, Sapporo, Hokkaido 001-0020, Japan*

*\* Corresponding authors: junzezhou@lbl.gov, afweber-bargioni@lbl.gov*

**Abstract**

Excitons in 2D material bubbles—nanoscale deformations in atomically thin materials, typically exhibiting a dome-like shape—are confined by the strain effect, exhibiting extraordinary emission properties, such as single photon generation, enhanced light emission, and spectrally tunable excitonic states. While the strain profiles of these bubbles have been extensively studied, this work provides an approach (1) to directly visualize the associated exciton properties, revealing an intrinsic emission wavelength shift of approximately 40 nm, and (2) actively modify local strain, enabling further exciton emission tuning over a range of 50 nm. These are achieved by emission mapping and nanoindentation using a dielectric near-field probe, which enables the detection of local emission spectra and emission lifetimes within individual bubbles. Statistical analysis of 67 bubbles uncovers an emission wavelength distribution centered around 780 nm. Furthermore, saturation behavior in the power-dependent studies and the associated lifetime change reveal the localized nature of the strain-induced states. These findings provide direct insights into the strain-localized emission dynamics in bubbles and establish a robust framework for non-destructive, reversible, and predictable nanoscale emission control, presenting a potential avenue for developing next-generation tunable quantum optical sources.

**Introduction**

Tunable single-photon emitters hold transformative potential for quantum information technologies, particularly in quantum information transduction.[1,2] Despite extensive research on single-photon sources, achieving tunability remains a significant challenge.[1,3,4] Nanobubbles (simply referred to as "bubbles" henceforth) in 2D Transition Metal Dichalcogenides (TMDs) are an emerging class of unconventional single-photon emitters [5] that are promising due to their strain-dependent optical properties and ability to maintain high emission quality at room temperature. Unlike the pillar-based approach,[6–8] which induces static strain through draping monolayers over lithographically-patterned substrate pillars, bubbles enable dynamic, in-situ strain control via mechanical manipulation, such as internal or external pressure. This enables real-time modifications to their shape, size, and strain profile, thereby tuning their emission properties.[9–12] This inherent flexibility distinguishes them from conventional single-photon sources and highlights their potential for tunable photonic applications.

The exceptional strain sensitivity of excitons in TMDs underpins the tunability of these bubble emitters. Their high exciton binding energies[13] and strong light–matter interactions enable stable photon emission,[14–16] while the intrinsic elasticity of TMDs supports significant strain modulation without structural degradation.[17–19] Localized strain engineering provides an effective strategy to tailor excitonic states and the electronic band structure,[9,20,21] offering precise spatial and spectral control of single-photon emission. Among various strain-inducing methods (such as pre-patterned substrates,[8,22] atomic force microscope (AFM) nanoindentation,[23,24] and engineered bubbles [5,25]), naturally formed bubbles during sample stacking present unique advantages, including predictable and well-defined strain profiles. These typically range in size from a few hundred nanometers to the micron scale.[5,11] According to elastic theory,[11,26,27] the strain in bubbles is proportional to the square of the aspect ratio *h*/*r*, where *h* is the maximum height and *r* is the radius. The balance between the 2D material–substrate adhesion energy and the material's in-plane stiffness determines the strain distribution, enabling tunable bubble emission through substrate engineering.[9]

Despite these advancements, key challenges remain in fully leveraging strain-engineered bubbles for deterministic, reversible tuning of excitonic emission energies. Conventional optical microscopes lack spatial resolution and the height/strain information to uncover the emission characteristics of the submicron-sized bubbles, whereas metallic near-field probes significantly reduce the emission upon contact, limiting their effectiveness for precise optical measurement.[20,28] Addressing these challenges is critical for bridging fundamental strain-induced band structure modulation with practical nanoscale control of light-matter interactions, including the ability to develop strain-tunable single-photon emitters.

In this study, we overcome these challenges by leveraging bubbles as a platform for strain engineering and employing a dielectric near-field tip to modulate strain. Using hyperspectral photoluminescence (PL) mapping, we first resolve the spatial and spectral emission profiles of strain-induced excitonic states, identifying a centralized wavelength distribution around 782 nm. By applying tip-induced strain, we achieve deterministic, reversible, and linear tuning of excitonic emission wavelengths, with shifts of up to 90 nm (~180 meV) from the unstrained exciton wavelength (energy). Furthermore, our results suggest that the localized strain acts as an exciton funnel, concentrating carriers within the strained region. Power-dependent studies on emission intensity and lifetime corroborate this interpretation, providing novel insights into the interplay between strain and excitonic dynamics. This work establishes a robust framework for strain engineering in TMDs and highlights the potential of bubbles as tunable quantum optical sources. Our findings offer potential implications for next-generation tunable single-photon emitters based on 2D materials, addressing a critical need in quantum information science.

**Results**

The bubbles are formed at the interface between the WSe$_2$ monolayer and hexagonal boron nitride (h-BN) (See Experimental section for fabrication details). To analyze the localized strain and the associated excitonic properties at the bubble site we used a dielectric near-field probe that is capable of simultaneously recording height information and PL signal.[29] During hyperspectral mappings, both excitation and collection were mediated via the fiber probe, as shown in **Fig. 1a** (See Experimental section for more details). The probe offers three key advantages: (i) State-of-art AFM-level topographical mapping, (ii) optical spectroscopy combined with a spatial resolution near the diffraction limit (approximately 300 nm for light at a wavelength of 750 nm),[29] suitable for bubbles larger than this dimension[9,11,26], (iii) and the ability to perform nanoindentation without quenching effects, thanks to the dielectric probe, enabling precise in-situ monitoring of the strain-induced excitonic emission.

By correlating the height and PL mappings on the bubbles, we analyze how the PL emission is influenced by their geometry, characterized by the height *(h)* and radius *(r)*, as illustrated in the schematic view in **Fig. 1a**. The height map in **Fig. 1b** shows that the bubble has a height *h* of approximately 40 nm and an effective width of approximately 350 nm, resulting in a low aspect ratio (*h/r* <0.2). Despite this low aspect ratio, the excitonic properties at the bubble site are notably altered. As shown in the correlated PL map and spectra (**Fig. 1c**), the bubble shows a threefold stronger emission peak compared to the one recorded from the flat region, accompanied by a wavelength shift of approximately 40 nm. This enhanced emission rate at the bubble site has been attributed to increased confinement in the strain-induced potential well, which improves the oscillator strength of excitons.[30] The localized strain can be estimated to be approximately 0.9%, based on the relation $\Delta E = \alpha \varepsilon$, where $\alpha$ is a constant representing the energy shift ($\Delta E$) per unit strain ($\varepsilon$), which is approximately 100 meV per percent of strain for the WSe$_2$ monolayer.[12,31,32] This strain estimate aligns with the calculated strain using the Föppl–von Kármán (FvK) equations (See **Fig. S1**).[27]

During the mapping experiment, an excitation power of 1.845 µW is used to ensure a good signal-to-noise ratio. Under this power, a small peak near the free exciton wavelength is observed alongside the strain-localized peak (**Fig. 1c**). However, this small peak disappears when excitation power is lowered to 0.05 µW as shown in **Fig. 1d**, suggesting that strain-localized states are reaching their maximum occupancy under the high power excitation, resulting in the emission of free excitonic states. This phenomenon is analogous to the saturation of defect-related emission in ZnO where, as the photoexcitation fluence and exciton density exceed the number density of the defect-related states, the band-edge emission increases.[33,34] Further evidence of the saturation effect will be provided by power-dependence measurements in the following section.

Notably, the strain-localized peak can be well fitted with a Lorentzian peak, indicating minimal inhomogeneous broadening. This suggests that the emission signal originates from a relatively homogenous region. The flatness of the bubble's flat top (see line profile in **Fig. 4d**) exhibits minimal strain fluctuations, as observed in the calculated strain map (**Fig. S1**). This consistence supports the notation that the flat top contributes to the uniformity of the emission signal in this region.

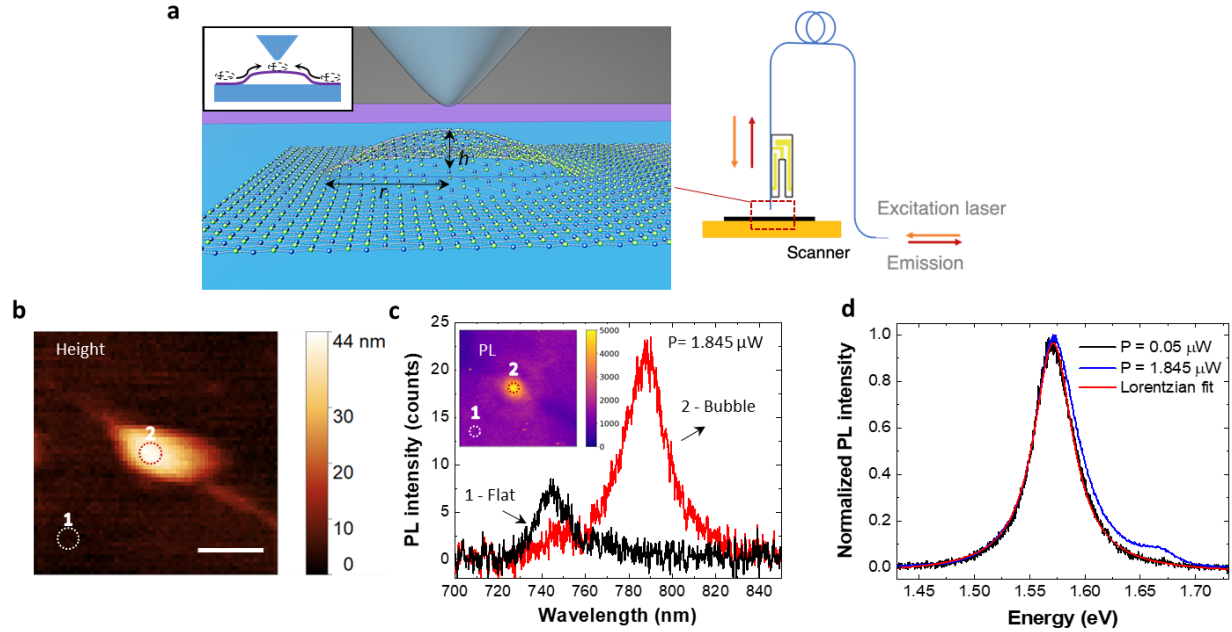

**Figure 1**. Simultaneous shear-force height and PL emission mapping of strained bubbles in WSe$_2$ monolayer with a dielectric scanning probe. (a) Schematic of the scanning optical microscope setup based on a fiber probe, where the optical path is in reflection mode—the excitation laser and emission signal are all coupled through the probe. The right graph shows a close-up view of the probe apex placed above the bubble at the interface between a WSe$_2$ and h-BN substrate. The geometry of the bubble is characterized by the height $h$ and its radius $r$. The inset represents the enhanced exciton density localized at the strained bubble. (b) Shear-force height map. The scale bar is 500 nm (c) PL spectra of the flat region (unstrained) and bubble site (strained) at areas 1 and 2 marked in (b). The excitation power (P) for the mapping is 1.845 µW. The inset image is the integrated PL intensity map recorded simultaneously with the one in (b). (d) PL spectra recorded with different excitation powers, where the spectrum at lower excitation power (P = 0.05 µW) is fitted with a single Lorentzian peak. The $x$-axis is converted from wavelength to energy for fitting purposes.

To statistically analyze the peak shift of bubble emissions in the sample stacked on h-BN, we perform large-scale hyperspectral scans. **Fig. 2a** presents a 20 × 30 µm$^2$ area PL map, composited from six separate scans acquired using the same probe. The light-emitting regions, with signal integrated over 700 – 850 nm at each pixel, correspond to TMD-covered areas, featuring luminescent hotspots from randomly distributed bubble formations (See **Fig. S2a** in the SI for height scans). The peak wavelengths of 67 bubbles (marked by the white circles) are analyzed, and **Fig. 2b** presents the histogram of their wavelength positions. The shifted wavelengths range from 760 nm to 800 nm and follow a Gaussian distribution centered at around 782 nm, corresponding to an 80 meV energy shift from the unstrained free excitonic energy.

To further understand the origin of the PL peak shift, we examine the geometric characteristics of the bubbles, particularly their aspect ratio ($h/r$), which is a critical parameter influencing their strain profiles. Statistical analysis of the height map (See **Fig. S2b** in the SI) reveals that the aspect ratio distribution is centered around 0.12, consistent with previous studies on the bubbles formed at the WSe$_2$–h-BN interface.[35] This is further confirmed by the analysis of a larger number of bubbles (See **Fig. S2c** in the SI). While the central wavelength shift aligns with the central aspect ratio distribution, exceptional cases – such as the tent-shaped bubble shown in **Fig. S3a** – deviate from this trend due to their irregular geometry. These bubbles exhibit a larger strain gradient,[26] resulting in an emission peak at a larger wavelength shift (**Fig. S3b**). Nevertheless, the general strain characteristics of the bubbles, governed by the material properties and substrate adhesion, impose a limit on their emission tunability.

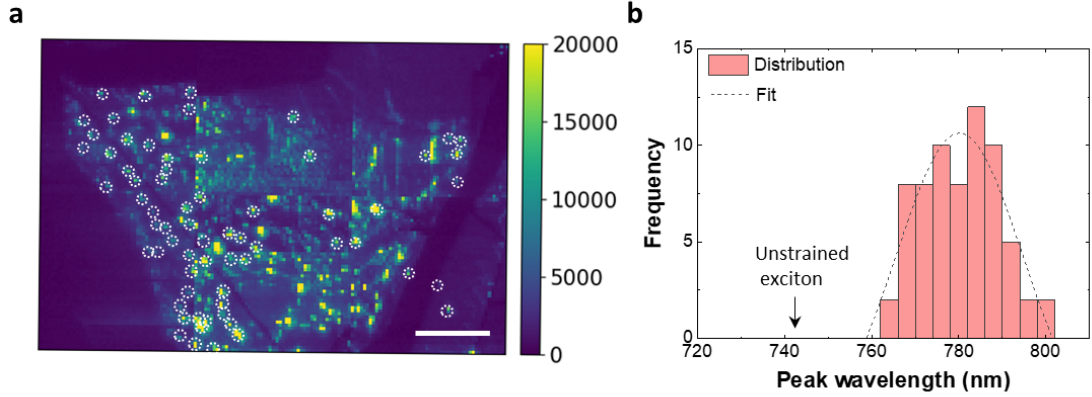

**Figure 2**. Universal PL characteristics within the sample. (a) PL emission map across an area of 20 × 30 μm². (b) Histogram of peak wavelength positions of the emission spectra from the 67 individual bubbles marked by the white circles. The distribution is fitted with a Gaussian curve.

To overcome the intrinsic limitations in the emission tunability of 2D bubbles, we utilize tip-induced strain to achieve deterministic control of the emission wavelength through localized nanoindentation. By performing nanoindentation and monitoring in-situ PL changes with the dielectric probe, we observe significant modifications in the emission spectrum, as illustrated in **Fig. 3a**. The evolution of the PL emission unfolds in three sequential stages. In the initial stage, as the tip approaches the bubble but does not make contact, the emission remains unchanged when the tip position changes from 8 nm to 0 nm. In the second stage, between 0 nm to approximately -7 nm, a slight blue shift in the emission is observed, indicating a reduction in strain. This shift can be attributed to a transition where the bubble's internal pressure is counteracted by the tip-induced pressure, causing the bubble to deform from outward bending to inward bending. The energy and intensity changes in this stage are minor, indicating that the applied strain is localized to a small contact region directly under the probe (where $r_{tip} \ll r_{bubble}$), while the overall bubble shape remains intact. This localized interaction aligns with the actual size of the probe[29] and nanoindentation experiments in the literature.[11] In the third stage, as the tip continues to indent the bubble, the emission spectrum exhibits "peak splitting," with the lowest energy peak undergoing a redshift. The maximum wavelength shift reaches approximately 832 nm, highlighting the significant impact of tip-induced strain on the bubble's excitonic properties.

In **Fig. 3b**, we compare the spectra recorded in different scenarios: flat region, bubble, and indented bubble. In addition to the indentation-induced emission peak, two extra peaks are observed, aligning with the emission positions at the free exciton and bubble emission energies. The appearance of the free exciton peak can be attributed to the saturation of the localized states, as an excitation power of 1.845 μW is used—similar to hyperspectral mapping—to minimize the stage drift while ensuring a good signal-to-noise ratio. The coexistence of the bubble emission peak highlights the localized nature of the nanoindentation, confined to the small area directly under the probe.[11] Notably, the majority of excitons emit from the lowest energy, which shifts as a function of the tip-induced strain. This enhanced emission is attributed to an additional exciton funneling channel created by the tip, which represents the lowest energy state. Consequently, exciton preferentially populates this lowest energy state, as illustrated in the schematics. In addition to the wavelength position shift, we also observe a local intensity maximum at the position $z = -29$ nm as marked by the black dotted circle in **Fig. 3a**. This observation aligns with recent studies suggesting a resonant condition can be achieved when the bent state hybridizes with the defect states.[12]

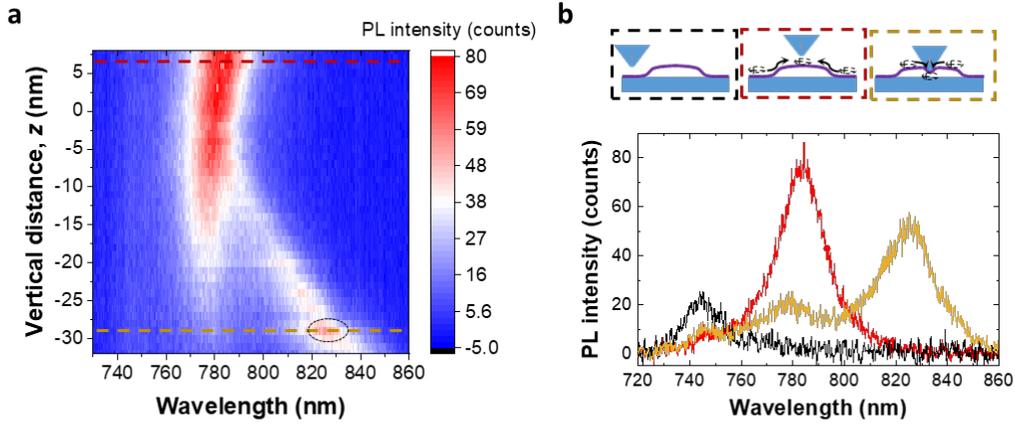

**Figure 3**. Evolution of PL spectra under nanoindentation. (a) PL spectra recorded as stage position changes from -32 nm to 8 nm (the initial scanning probe–sample distance). (b) Emission spectra recorded in three different cases: Tip on the flat region, tip on the bubble at $z = 6$ nm (marked by the red dashed line in (a)), and tip on the intended bubble at $z = -29$ nm (marked by the yellow dashed line in (a)).

In the zoomed-in view of PL evolution (**Fig. 4a**), we observe a linear wavelength shift as a function of tip–sample distance. This reproducible linearity (See extra data in **Fig. S4** in the SI for the reproducibility) can be interpreted using the small nanoindentation approximation, where the strain is proportional to $\delta/r$, with $\delta$ representing the indentation depth and $r$ the bubble radius. This approximation is valid in the elastic deformation regime, where the materials' response is proportional to the applied force ($F$). The linearity of the wavelength shift implies that the tunability is not only controllable but also predictable by adjusting the indentation depth through the vertical movement of the sample stage. Additionally, since $F$ is proportional to the amplitude damping of the tuning fork oscillation, the indentation depth can also be controlled by fine-tuning the tuning fork setpoint, which in turn adjusts the tip-induced strain. This approach is applied in the following section to obtain static PL spectra under tip strain (inset of **Fig. 5f**). Such precise control is invaluable for practical applications where specific emission wavelengths are required.

The elastic nature of this response is further confirmed by the unchanged bubble shape before and after the nanoindentation, as shown in **Fig. 4a** and **4d**. The stability of the bubble's geometry under repeated indentations indicates that the material remains within its elastic limit, avoiding any plastic or permanent deformation. As a result, the intrinsic emission of the bubble remains consistent, and the strain-induced shifts in the emission wavelength are fully reversible when the tip is retracted, as demonstrated by the overlapped PL spectra before and after indentation in **Fig. 4c**. This ability to non-destructively induce, linearly control, and reversibly tune localized emission is essential for practical applications such as strain-induced quantum emitters.

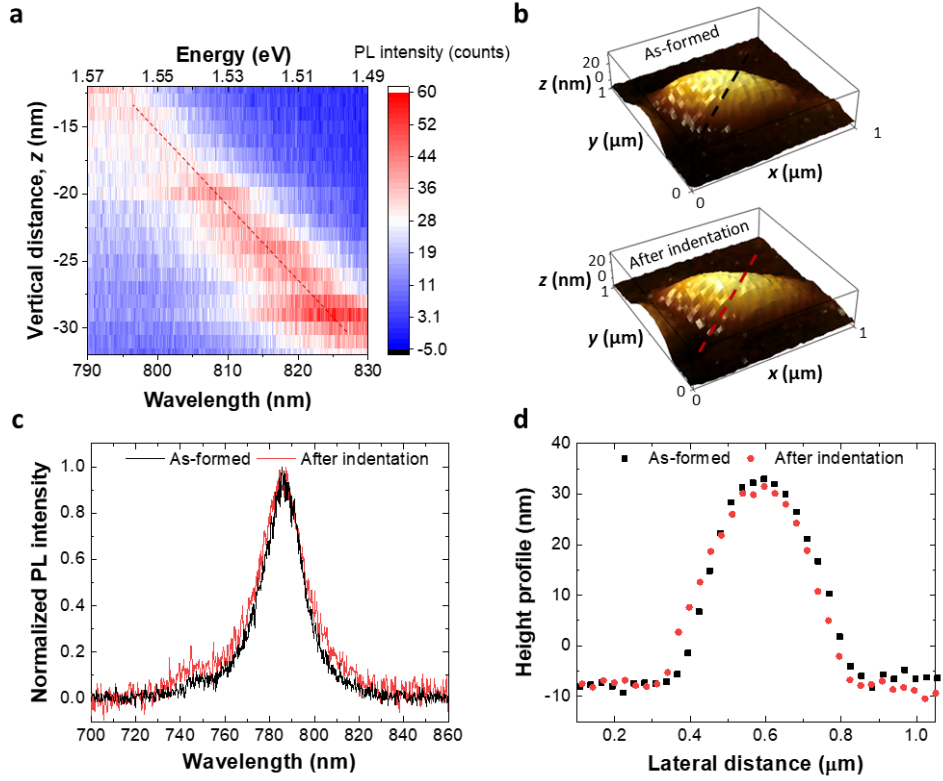

**Figure 4**. Elastic deformation induced by the tip. (a) Linear evolution of the emission wavelength as a function of nanoindentation depth. (b, c, d) Topography, emission spectra, and height profiles of the bubble before and after the nanoindentation.

.

## Discussion

The energy band diagram for the indented bubble is proposed in **Fig. 5a, 5b, and 5c**. Compared to the cases where the probe scans over the flat region or unindented bubbles, nanoindentation introduces an additional localized state directly under the tip. This state enables excitons to be trapped and guided into the lowest energy configuration within the indented region. As previously discussed, high excitation power is required in both hyperspectral scans and nanoindentation experiments to minimize stage drift. Under these conditions, the strain-localized states reach saturation, where exciton density exceeds the maximum capacity, preventing additional excitons from remaining confined to the lowest energy state. This saturation results in the appearance of both free exciton and bubble emission peaks in the nanoindented PL spectrum.

To validate the proposed mechanism, we investigate the exciton dynamic in the localized state by conducting power-dependent studies and analyzing the associated lifetime variations. The PL lifetimes were extracted by fitting the time-resolved PL data using a single-exponential decay model, $I(t) = I_0 exp(-t/\tau_{PL})$, where $I(t)$ is the intensity as a function of time, $I_0$ is the initial intensity, and $\tau_{PL}$ is the PL lifetime. This fitting approach is applied consistently across all measured regions to ensure comparability. The fitting results reveal distinct differences between free exciton and bubble emission. As shown in **Fig. 5d** and **5g**, the free exciton emission measured on the flat region remains in the linear region, with PL intensity increasing linearly with excitation power. The corresponding lifetime remains constant at approximately 652 ps, indicating minimal power dependence. In contrast, the bubble emission exhibits clear signs of saturation with increasing excitation power (**Fig. 5e**), described by the fit $I = I_{sat}P/(P + P_{Sat})$, where the $I_{sat}$ is the saturated emission intensity and $P_{Sat}$ is saturated power. The fitted $P_{Sat}$ is 0.82 µW, representing a low excitation power. Concurrently, the lifetime of bubble emission decreases from 832 ps to 376 ps with increasing power (**Fig. 5h**).

We argue that the decreased lifetime is mainly due to the increase in non-radiative recombination at the physically confined strain-localized state. At low excitation power, the lifetime is longer than the free exciton lifetime. This extended lifetime is expected because strain-localized states typically exhibit lower non-radiative recombination rates, such as reduced phonon scattering.[32] As the excitation increases, the PL lifetime decreases. This behavior occurs because the photoexcited carrier in the confined 2D exciton reservoir saturates at the bubble site, as confirmed by the intensity saturation, which enhances the likelihood of non-radiative processes such as exciton–exciton annihilation[36] or auger recombination[37]. These processes introduce additional decay pathways that compete with radiative recombination, thereby shortening the $\tau_{PL}$. Additionally, although the indenter is purely dielectric, the higher refractive index ($n$ ~1.5)[38] increases the local density of photonic states, which can enhance radiative recombination and further influence the measured lifetime. Notably, we have deterministically confirmed the longer lifetime of strain-localized states using our probe, which aligns with previous reports of bubble-related single photon emissions at low temperatures, where lifetimes typically reach the nanosecond range.[6,39–41] However, it is not possible to lower excitation power further to explore this behavior more deeply in our current setup due to detection limits at ambient conditions.

Building on these findings, we measure the lifetime of the tip-strain-induced emission line by reducing the tuning fork's free amplitude setpoint from 95% to 64%. The maximum PL emission shifts by around 43 nm as shown in **Fig. 5f**. The lifetime measurement is performed under an excitation power of 0.4235 μW with a 1-second integration time. The measured lifetime of the 825 nm line is 414 ps, shorter than that of the un-indented bubble emission line (527 ps), as shown in **Fig. 5i**. This shorter lifetime indicates faster exciton dynamics in the tip-strain-induced state, likely due to the increased localization and lower exciton occupancy density within the more physically confined strained area. Due to instrumental limitations, we are unable to perform a stable power-dependent study under this configuration. Nevertheless, our analysis of bubble strain emission strongly suggests that the tip-induced state also reaches saturation under high excitation conditions.

To contextualize these findings, we present schematic diagrams illustrating the mechanisms of excitonic transitions observed in static PL spectra in **Fig. 3b**. For the bubble emission spectrum, the strain-localized state accounts for the main peak, while the free exciton peak arises due to the saturation, as illustrated in **Fig. 5b**. Similarly, in the tip-strain scenario, the emission spectrum comprises three distinct transitions. Among these, the transition associated with the tip-strain state dominates, as it represents the lowest energy state, as illustrated in **Fig. 5c**. These observations highlight the interplay between localized states and free excitons under varying strain conditions and demonstrate the tunability of excitonic emission in 2D materials controlled through the tip strain.

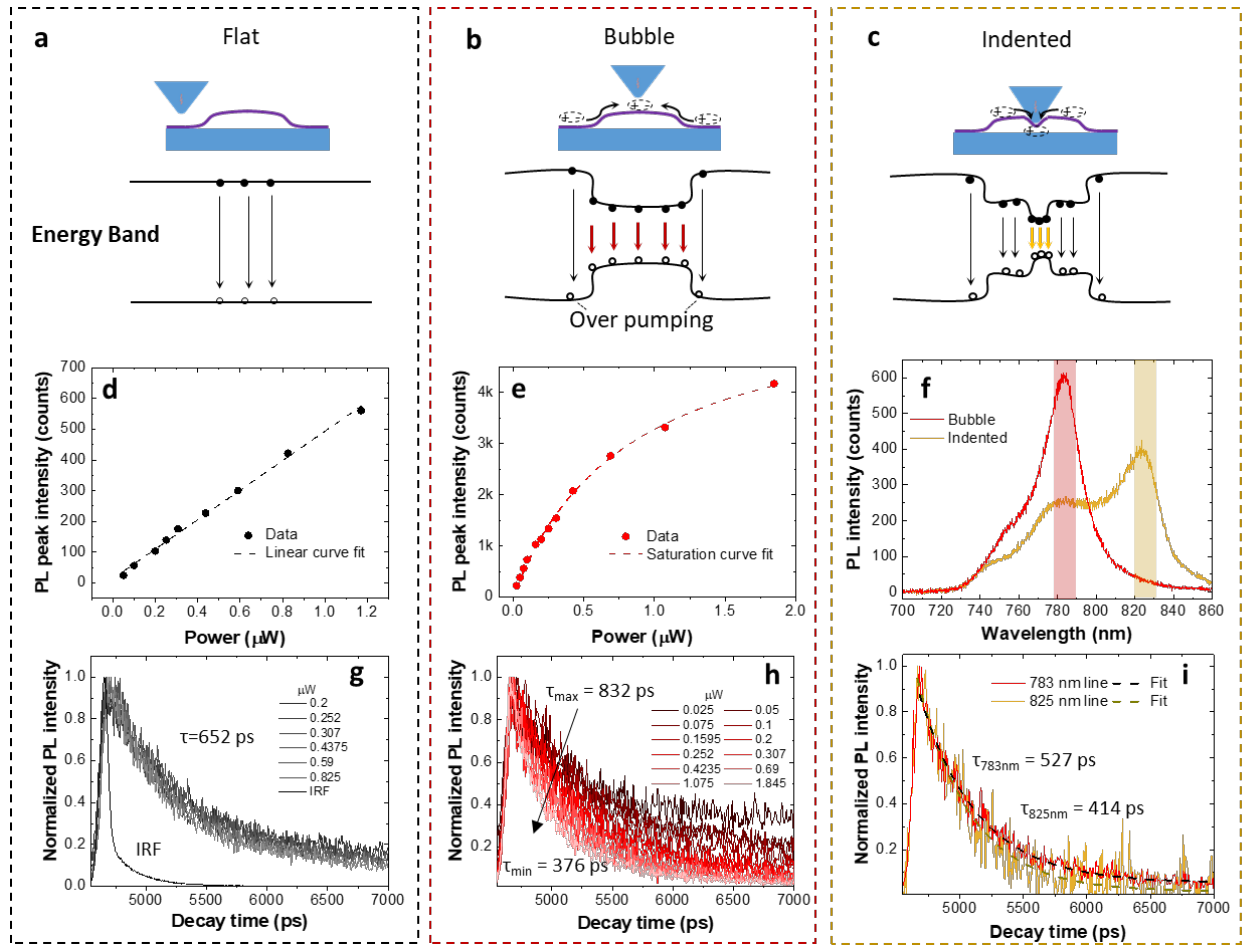

**Figure 5**. Photophysics of the strain-induced localized states. (a, b, c) Energy band diagrams and proposed mechanism of the electron transitions in three different cases: flat region, bubble with built-in strain, and bubble under tip-induced strain. (d, e) Power-dependent PL intensity measured on the flat region and bubble, and the fittings to a linear curve and a saturation curve, respectively. (g, h) PL intensity decay as a function of excitation power measured on the flat region and bubble with the filtered peak at 743 nm and 780 nm, respectively. (f) PL spectra recorded when the tuning fork feedback is set at the setpoints of 95% and 64%. (i) PL intensity decays of the bubble emission and the tip-strain-induced peaks (wavelength ranges outlined by red and yellow regions in (g). The excitation power is 0.4235 μW.

## Conclusion

In summary, we investigate the excitonic properties induced by the built-in strain in 2D material bubbles and demonstrate their tunability using a dielectric near-field probe. By leveraging the capability of our probe to perform simultaneous topographic and spectroscopic measurements, we achieved a high-resolution spectral readout at the bubble sites. This allows us to reveal a centralized peak shift of approximately 80 meV, determined by the intrinsic properties of the substrate and 2D material. Furthermore, we demonstrate a reversible and precise linear tunability of strain-localized states through tip nanoindentation, achieving an additional energy shift of up to 180 meV. Our investigations into the exciton dynamics reveal that exciton emission at the strained bubble exhibits a longer lifetime compared to free exciton states. The saturation of these states at low excitation power highlights their efficiency in excitonic trapping and recombination processes. Unlike prior methods, which often suffered from spatial mismatch or introduced charging effects, our approach provides a robust, non-destructive, and repeatable means to explore strain-tunable excitonic properties with nanoscale precision. These findings highlight the potential of straining engineering as a powerful tool for enabling tuning the excitonic properties in 2D materials. The integration of our dielectric probe-based technique with other platforms such as a cryogenic chamber could further broaden its applicability, paving the way for the development of next-generation quantum emitters.

## Experimental Section

**Sample preparation**: The $WSe_2$ monolayer and hexagonal boron nitride (h-BN) are mechanically exfoliated. Bubbles are created by mechanically stacking the $WSe_2$ monolayer on h-BN. During this process, bubbles naturally form as interfacial contaminants (primarily water and hydrocarbons) are trapped and compressed by the strong adhesive forces between 2D materials and the substrate.[11,25] In the conventional transfer method, h-BN is exfoliated and transferred to a substrate first, followed by exfoliating and transferring the TMD monolayer onto the h-BN-coated substrate. To promote the formation of uniformly shaped bubbles, we modify this approach by directly exfoliating the h-BN onto the substrate before transferring the TMD monolayer. This ensures a cleaner and flatter surface, supporting consistent adhesion.

**The nanoimprinted near-field fiber probe**: (1) A cleaved single-mode fiber (630HP, Thorlabs) is dip-coated with Ormocomp photoresist. (2) The collated fiber is mounted onto a piezo stage above the transparent mold featuring an inverted pyramidal structure. (3) The fiber core is located by coupling red light through the fiber, and the transmitted light is monitored via a camera. This enables precise alignment of the fiber core with the center of the mold. (4) UV light is illuminated from below the mold to expose and cure the Ormocomp for ~3 min, solidifying the pyramid structure. (5) The mold is carefully removed, leaving a well-defined pyramid shape integrated onto the fiber facet, with the pyramid's base covering the fiber core.

**Steady-state PL and time-resolved PL measurement**: The PL measurements are carried out in a fiber-in–fiber-out configuration. In this setup, a 633 nm excitation laser (FWHM= 6 ps, repetition rate of 60 MHz), derived from the Supercontinuum laser source (Leukos LMS 7026), is modulated using an acoustic-optics modulator and coupled to the fiber using the Thorlabs MBT612D fiber launch stage with a 10× objective lens (0.25 NA). The emitted signal from beneath the probe is collected through the same path and directed to the spectrometer (Andor Kymera 328i), which records a spectrum ranging from 624 nm to 913 nm. The lifetime is measured using a single photon avalanche diode (PicoQuant PDM series) and a single photon counting module (HydraHarp 400). The time resolution of this setup is ~60 ps.

**Nanoimaging and nanoindentation**: The scanning near-field optical microscope is based on the Horiba CombiScope microscope equipped with a shear-force head. The near-field probe is attached to a tuning fork oscillating at around 35 kHz, which modulates the tip–sample distance. Typically, the tuning fork oscillation amplitude is set to 95% of its free amplitude, maintaining a constant tip-to-sample distance. During scanning, the tip records both the shear-force height information and the emission spectra simultaneously at each pixel. A typical integration time of 0.1 s per pixel is used for PL data acquisition. Two approaches are employed in the nanoindentation experiments: (1) The spectra stack function in the Horiba software is used to capture sequential PL spectra while varying the stage height incrementally with a step size of 1 nm and an integration of 0.1 s. (2) The setpoint of the tuning fork is reduced to achieve a fixed indentation depth, allowing for the acquisition of spectra with a longer integration time.

# Reference

[1] Y. Yu, I. C. Seo, M. Luo, K. Lu, B. Son, J. K. Tan, D. Nam, *Nanophotonics* **2024**, *13*, 3615.
[2] A. Alfieri, S. B. Anantharaman, H. Zhang, D. Jariwala, *Adv. Mater.* **2023**, *35*, e2109621.
[3] I. Aharonovich, D. Englund, M. Toth, *Nat. Photonics* **2016**, *10*, 631.
[4] G. Zhang, Y. Cheng, J.-P. Chou, A. Gali, *Appl. Phys. Rev.* **2020**, *7*, DOI 10.1063/5.0006075.
[5] G. D. Shepard, O. A. Ajayi, X. Li, X.-Y. Zhu, J. Hone, S. Strauf, *2D Mater.* **2017**, *4*, 021019.
[6] C. Palacios-Berraquero, D. M. Kara, A. R.-P. Montblanch, M. Barbone, P. Latawiec, D. Yoon, A. K. Ott, M. Loncar, A. C. Ferrari, M. Atatüre, *Nat. Commun.* **2017**, *8*, 15093.
[7] H. Zhao, M. T. Pettes, Y. Zheng, H. Htoon, *Nat. Commun.* **2021**, *12*, 6753.
[8] K. Parto, S. I. Azzam, K. Banerjee, G. Moody, *Nat. Commun.* **2021**, *12*, 3585.
[9] A. V. Tyurnina, D. A. Bandurin, E. Khestanova, V. G. Kravets, M. Koperski, F. Guinea, A. N. Grigorenko, A. K. Geim, I. V. Grigorieva, *ACS Photonics* **2019**, *6*, 516.
[10] D. Zhang, L. Gan, J. Zhang, R. Zhang, Z. Wang, J. Feng, H. Sun, C.-Z. Ning, *ACS Nano* **2020**, *14*, 6931.
[11] E. Khestanova, F. Guinea, L. Fumagalli, A. K. Geim, I. V. Grigorieva, *Nat. Commun.* **2016**, *7*, 12587.
[12] P. Hernández López, S. Heeg, C. Schattauer, S. Kovalchuk, A. Kumar, D. J. Bock, J. N. Kirchhof, B. Höfer, K. Greben, D. Yagodkin, L. Linhart, F. Libisch, K. I. Bolotin, *Nat. Commun.* **2022**, *13*, 7691.
[13] A. Chernikov, T. C. Berkelbach, H. M. Hill, A. Rigosi, Y. Li, O. B. Aslan, D. R. Reichman, M. S. Hybertsen, T. F. Heinz, *Phys. Rev. Lett.* **2014**, *113*, 076802.
[14] K. F. Mak, J. Shan, *Nat. Photonics* **2016**, *10*, 216.
[15] S. Liu, A. Granados Del Águila, X. Liu, Y. Zhu, Y. Han, A. Chaturvedi, P. Gong, H. Yu, H. Zhang, W. Yao, Q. Xiong, *ACS Nano* **2020**, *14*, 9873.
[16] A. Srivastava, M. Sidler, A. V. Allain, D. S. Lembke, A. Kis, A. Imamoğlu, *Nat. Nanotechnol.* **2015**, *10*, 491.
[17] D. Akinwande, C. J. Brennan, J. S. Bunch, P. Egberts, J. R. Felts, H. Gao, R. Huang, J.-S. Kim, T. Li, Y. Li, K. M. Liechti, N. Lu, H. S. Park, E. J. Reed, P. Wang, B. I. Yakobson, T. Zhang, Y.-W. Zhang, Y. Zhou, Y. Zhu, *Extreme Mech. Lett.* **2017**, *13*, 42.
[18] Z. Li, Y. Lv, L. Ren, J. Li, L. Kong, Y. Zeng, Q. Tao, R. Wu, H. Ma, B. Zhao, D. Wang, W. Dang, K. Chen, L. Liao, X. Duan, X. Duan, Y. Liu, *Nat. Commun.* **2020**, *11*, 1151.
[19] P. Johari, V. B. Shenoy, *ACS Nano* **2012**, *6*, 5449.
[20] T. P. Darlington, C. Carmesin, M. Florian, E. Yanev, O. Ajayi, J. Ardelean, D. A. Rhodes, A. Ghiotto, A. Krayev, K. Watanabe, T. Taniguchi, J. W. Kysar, A. N. Pasupathy, J. C. Hone, F. Jahnke, N. J. Borys, P. J. Schuck, *Nat. Nanotechnol.* **2020**, *15*, 854.
[21] B. Aslan, M. Deng, T. F. Heinz, *Phys. Rev. B Condens. Matter* **2018**, *98*, DOI 10.1103/physrevb.98.115308.
[22] E. S. Yanev, T. P. Darlington, S. A. Ladyzhets, M. C. Strasbourg, C. Trovatello, S. Liu, D. A. Rhodes, K. Hall, A. Sinha, N. J. Borys, J. C. Hone, P. J. Schuck, *Nat. Commun.* **2024**, *15*, 1543.
[23] M. R. Rosenberger, C. K. Dass, H.-J. Chuang, S. V. Sivaram, K. M. McCreary, J. R. Hendrickson, B. T. Jonker, *ACS Nano* **2019**, *13*, 904.
[24] J.-P. So, H.-R. Kim, H. Baek, K.-Y. Jeong, H.-C. Lee, W. Huh, Y. S. Kim, K. Watanabe, T. Taniguchi, J. Kim, C.-H. Lee, H.-G. Park, *Sci. Adv.* **2021**, *7*, DOI 10.1126/sciadv.abj3176.
[25] C. Di Giorgio, E. Blundo, G. Pettinari, M. Felici, F. Bobba, A. Polimeni, *Adv. Mater. Interfaces* **2022**, *9*, 2102220.
[26] Z. Dai, Y. Hou, D. A. Sanchez, G. Wang, C. J. Brennan, Z. Zhang, L. Liu, N. Lu, *Phys. Rev. Lett.* **2018**, *121*, 266101.
[27] T. P. Darlington, A. Krayev, V. Venkatesh, R. Saxena, J. W. Kysar, N. J. Borys, D. Jariwala, P. J. Schuck, *J. Chem. Phys.* **2020**, *153*, 024702.
[28] K.-D. Park, O. Khatib, V. Kravtsov, G. Clark, X. Xu, M. B. Raschke, *Nano Lett.* **2016**, *16*, 2621.
[29] J. Zhou, A. Gashi, F. Riminucci, B. Chang, E. Barnard, S. Cabrini, A. Weber-Bargioni, A. Schwartzberg, K. Munechika, *Rev. Sci. Instrum.* **2023**, *94*, 033902.
[30] H. Moon, G. Grosso, C. Chakraborty, C. Peng, T. Taniguchi, K. Watanabe, D. Englund, *Nano Lett.* **2020**, *20*, 6791.
[31] R. Frisenda, M. Drüppel, R. Schmidt, S. Michaelis de Vasconcellos, D. Perez de Lara, R. Bratschitsch, M. Rohlfing, A. Castellanos-Gomez, *npj 2D Materials and Applications* **2017**, *1*, 1.
[32] P. Qi, Y. Luo, B. Shi, W. Li, D. Liu, L. Zheng, Z. Liu, Y. Hou, Z. Fang, *eLight* **2021**, *1*, DOI 10.1186/s43593-021-00006-8.
[33] J. G. Simmons Jr, J. V. Foreman, J. Liu, H. O. Everitt, *Appl. Phys. Lett.* **2013**, *103*, 201110.
[34] T. Nakamura, *Phys. Rev. Appl.* **2016**, *6*, DOI 10.1103/physrevapplied.6.044009.
[35] E. Blundo, T. Yildirim, G. Pettinari, A. Polimeni, *Phys. Rev. Lett.* **2021**, *127*, 046101.
[36] L. Yuan, L. Huang, *Nanoscale* **2015**, *7*, 7402.
[37] M. Kulig, J. Zipfel, P. Nagler, S. Blanter, C. Schüller, T. Korn, N. Paradiso, M. M. Glazov, A. Chernikov, *Phys. Rev. Lett.* **2018**, *120*, 207401.
[38] T. Gissibl, S. Wagner, J. Sykora, M. Schmid, H. Giessen, *Opt. Mater. Express* **2017**, *7*, 2293.
[39] S. Kumar, A. Kaczmarczyk, B. D. Gerardot, *Nano Lett.* **2015**, *15*, 7567.
[40] Y.-M. He, G. Clark, J. R. Schaibley, Y. He, M.-C. Chen, Y.-J. Wei, X. Ding, Q. Zhang, W. Yao, X. Xu, C.-Y. Lu, J.-W. Pan, *Nat. Nanotechnol.* **2015**, *10*, 497.
[41] L. Sortino, P. G. Zotev, C. L. Phillips, A. J. Brash, J. Cambiasso, E. Marensi, A. M. Fox, S. A. Maier, R. Sapienza, A. I. Tartakovskii, *Nat. Commun.* **2021**, *12*, 6063.